\documentclass[english,preprint]{aastex}
\shorttitle{}
\shortauthors{}

\begin{document}

\title{Magnetic Fields and Infall Motions in NGC 1333 IRAS 4}

\author{Michael Attard\altaffilmark{1}\email{mattard@uwo.ca} Martin Houde\altaffilmark{1},
Giles Novak\altaffilmark{2}, Hua-bai Li\altaffilmark{3}, John E.
Vaillancourt\altaffilmark{4}, C. Darren Dowell\altaffilmark{4,5}
Jacqueline Davidson\altaffilmark{6}, \and Hiroko Shinnaga\altaffilmark{7}}

\altaffiltext{1}{Department of Physics and Astronomy, The University of Western Ontario, London, ON N6A 3K7, Canada}

\altaffiltext{2}{Department of Physics and Astronomy, Northwestern University, Evanston, IL 60208}

\altaffiltext{3}{Harvard-Smithsonian Center for Astrophysics, MS-78, Cambridge, MA 02138}

\altaffiltext{4}{Division of Physics, Mathematics, \& Astronomy, California Institute of Technology, 1200 E. California Blvd., MS 320-47, Pasadena, CA 91125}

\altaffiltext{5}{Jet Propulsion Laboratory, California Institute of Technology, Pasadena, CA, 91109}

\altaffiltext{6}{School of Physics, The University of Western Australia, Crawley, WA 6009, Australia}

\altaffiltext{7}{Caltech Submillimeter Observatory, 111 Nowelo St., Hilo, HI 96720}
\begin{abstract}
We present single-dish $350$~$\mu m$ dust continuum polarimetry
as well as $\mathrm{HCN}$ and $\mathrm{HCO^{+}}$ $J=4\rightarrow3$
rotational emission spectra obtained on NGC 1333 IRAS 4. The polarimetry
indicates a uniform field morphology over a $20$\arcsec~radius
from the peak continuum flux of IRAS 4A, in agreement with models
of magnetically supported cloud collapse. The field morphology around
IRAS 4B appears to be quite distinct however, with indications of
depolarization observed towards the peak flux of this source. Inverse
P-Cygni profiles are observed in the $\mathrm{HCN\:}J=4\rightarrow3$
line spectra towards IRAS 4A, providing a clear indication of infall
gas motions. Taken together, the evidence gathered here appears to
support the scenario that IRAS 4A is a cloud core in a critical state
of support against gravitational collapse. 
\end{abstract}


\section{\label{sec:Introduction}Introduction}

The process of star formation is a key phenomenon in astrophysics.
It touches upon a diverse range of topics including galactic evolution,
stellar evolution, planet formation, and astrobiology. Yet despite
the fundamental nature of this process, it remains poorly understood
\citep{Crutcher 2008}. A key issue is determining the support mechanism(s)
that governs star formation, as it is clear stars cannot be forming
at a free fall timescale \citep{Shu 1987}. Two competing ideas currently
strive to explain the support mechanism; magnetism plus ambipolar
diffusion and turbulence plus a weak magnetic field \citep{McKee 2007}.
At present, a consensus on the nature of the support mechanism has
not been realized.

A particularly important quantity for star formation theory is the
mass-to-flux ratio, $M/\Phi$, where $\Phi$ is the magnetic flux
within a region enclosing a mass $M$ \citep{Nakamura 2008}. Magnetic
support models predict the critical value of this ratio to be equal
to $c_{\Phi}/\sqrt{G}$ \citep{Mouschovias 1976}, where $G$ is the
gravitational constant and $c_{\Phi}=0.12$ \citep{Tomisaka 1988}.
It follows that the value of $M/\Phi$ normalized to $c_{\Phi}/\sqrt{G}$
should be close to unity in regions near the point of collapse. Turbulent
models place no such restrictions on the mass-to-flux ratio but do
predict a chaotic field morphology assuming that flux-freezing holds.
The mass-to-flux ratio can be determined from observables and thus
may be used as an important indicator as to the nature of the support
mechanism. 

At a distance of $\approx300\:\mathrm{pc}$ %
\footnote{The distance of $300$ pc will be assumed throughout this work.%
} \citep{Girart 2006}, a known site of clustered low and intermediate
mass star formation, and possessing young embedded cores at an age
of $\approx1\:\mathrm{Myr}$ \citep{Hatchell 2005}, NGC 1333 is an
ideal target for studying the onset of clustered star formation. Two
such embedded cores of interest include NGC 1333 IRAS 4A ($\alpha_{2000}$
= 3h29m10.42s, $\delta_{2000}$ = $+31^{\circ}13$\arcmin$35.4$\arcsec,
henceforth 4A) and NGC 1333 IRAS 4B ($\alpha_{2000}$ = 3h29m12.06s,
$\delta_{2000}$ = $+31^{\circ}13$\arcmin$10.8$\arcsec, henceforth
4B). Note that $\alpha_{2000}$ and $\delta_{2000}$ will denote J2000
epoch Right Ascension and Declination coordinates throughout this
paper, respectively ($\alpha_{o}$ and $\delta_{o}$ will denote offsets
in Right Ascension and Declination from a particular reference point,
respectively). The source 4A has been extensively studied in the past
\citep{Sandell 1991, DiFrancesco 2001, Girart 2006}. We have carried
out continuum polarization and spectroscopic observations using, respectively,
SHARP at $350\:\mathrm{\mu m}$ towards 4A/4B and the $300-400$$\mathrm{GHz}$
heterodyne receiver at the Caltech Submillimeter Observatory (CSO)
on 4A. It is these data and the subsequent analysis that are presented. 

Previous work has greatly improved our knowledge of both 4A and 4B.
The interferometric detection of inverse P-Cygni profiles towards
both of these sources provides strong evidence for infalling gas motions
(Di Francesco et al. 2001, hereafter DF2001). The detections were
made in $\mathrm{H_{2}CO}$ $3_{21}\rightarrow2_{11}$ emission and
have allowed the determination of infall speeds of $0.68$ and $0.47\:\mathrm{km/s}$
for 4A and 4B, respectively. DF2001 also provide simple mass estimates
for the gas in each source and find $0.71$ $\mathrm{M_{\odot}}$
and $0.23$ $\mathrm{M_{\odot}}$ within corresponding radii of $9$\arcsec
($0.013\:\mathrm{pc}$) and $6$\arcsec ($0.009\:\mathrm{pc}$) from
the peak flux of 4A and 4B, respectively. Finally, mass accretion
rates of $1.1\times10^{-4}$ $\mathrm{M_{\odot}/yr}$ and $3.7\times10^{-5}$
$\mathrm{M_{\odot}/yr}$ were calculated for both 4A and 4B, respectively.

Recent $877\mathrm{\:\mu m}$ continuum polarimetry done at the Submillimeter
Array (SMA) has indicated the presence of a well defined pinch in
the magnetic field morphology around 4A (Girart et al. 2006, hereafter
G2006). From these measurements the authors estimate the mass-to-flux
ratio to be $\approx1.7$ times the critical value of collapse. Along
with their computation of the ratio of the turbulent to magnetic energy,
$\beta_{\mathrm{turb}}\equiv\sigma_{\mathrm{turb}}^{2}/V_{A}^{2}\approx8\times10^{-4}\left(\delta\theta_{\mathrm{int}}/1^{\circ}\right)^{2}=0.02$
(where $\sigma_{\mathrm{turb}}$ is the turbulent line width, $V_{A}$
is the Alfv$\acute{\mathrm{e}}$n speed, and $\delta\theta_{\mathrm{int}}$
is the intrinsic dispersion in the polarization vectors; Lai et al.
2002), the authors conclude that 4A is an example of a magnetically
dominated collapsing cloud core. A cloud mass of $1.2\,\mathrm{M_{\odot}}$
within a radial distance of $3$\arcsec ($0.004\,\mathrm{pc}$) from
the peak flux of 4A is also traced by their dust continuum measurements.
Taken together, the DF2001 \& G2006 results present compelling evidence
for the notion that the physics of 4A is consistent with standard
magnetically-regulated star formation theory.

One problem not addressed in these works is the variation of physical
parameters as a function of spatial scales. Of particular interest
here is the variation in the magnetic field morphology with spatial
scales; models predict that regions undergoing collapse will drag
in the field lines towards the central condensation producing an hourglass
morphology. Further out from the condensation, the field morphology
should remain in its ambient state \citep{Fielder 1993}. The results
presented in G2006 illustrate an example of this hourglass morphology
at a resolution of $\approx1$\arcsec ($0.001\:\mathrm{pc}$). However,
due to this small spatial scale G2006 was unable to sample the field
morphology at scales larger than $\approx10$\arcsec ($0.015\:\mathrm{pc}$)
where models predict the field to be uniform. Similarly for the spectroscopy
work, the high resolution attained through interferometry by DF2001
($\approx2$\arcsec or $0.003\:\mathrm{pc}$) allowed them to identify
infall signatures out to a distance of $\approx4$\arcsec ($0.006\:\mathrm{pc}$)
from the peak flux of 4A and 4B. It is not clear from their results
whether or not infall motions are occurring further out from the peak
positions. 

The aim of this study is to address the problem of spatial scale variation
in magnetic fields and infall motions and to complement the work of
G2001 and DF2001. This will be done by analyzing single-dish observations
obtained at larger spatial scales and comparing these results with
the aforementioned papers. In the following sections we describe both
polarimetric and spectroscopic observations carried out at the CSO
and discuss the implications of our findings. In Section \ref{sec:2}
we discuss our observations. Section \ref{sec:3} will cover a general
discussion and analysis of our results, and finally in Section \ref{sec:4}
we state our conclusions.

\section{\label{sec:2}Observations}

The following two sub-sections describe the submillimeter dust continuum
polarimetry and spectroscopic observations acquired by our group.
In Section \ref{sub:Polarimetry} we discuss $350\:\mu m$ polarimetry
data collected in September 2008 using SHARP, the SHARC-II polarimeter.
Section \ref{sub:Spectroscopy} will discuss the $\mathrm{HCN}$~$J=4\rightarrow3$
spectroscopic observations taken with the CSO $300-400$ $\mathrm{GHz}$
heterodyne receiver in September 2000.

\subsection{\label{sub:Polarimetry}Polarimetry}

Dust continuum polarimetry was done with SHARP at $350\:\mathrm{\mu m}$
with a spatial resolution of $\approx10$\arcsec ($0.015\:\mathrm{pc}$).
SHARP is a fore-optics addition to the SHARC-II camera that enables
this instrument to be used as a sensitive polarimeter \citep{Li 2008}.
Although both 4A and 4B are studied here, the telescope was pointed
onto 4B so as to provide the best possible chance of detecting polarization
on this fainter source. Our map was calibrated with approximately
3 hours worth of data obtained on W3(OH) during the same observing
run. In order to account for any random or systematic uncertainties
that may remain after applying our standard data reduction pipeline,
a reduced-$\chi^{2}$ analysis was performed on the data. The measurement
uncertainty was correspondingly inflated, on a pixel-by-pixel basis,
such that the reduced-$\chi^{2}=1$. Our results are shown in Figure
1 and represent approximately 10.5 hrs of observing time.

Several points are worth mentioning from Figure 1: (1) the extended
magnetic field around 4A is clearly sampled in our map. The nature
of the extended field appears to be uniform out to a distance of $\approx20$\arcsec
($0.03\:\mathrm{pc}$) from the peak of this source. This is consistent
with the $850\:\mathrm{\mu m}$, effective $20$\arcsec~beamwidth,
SCUPOL observations taken at the JCMT \citep{Matthews 2009}. Our
data enables a rough upper-limit to be placed on the size of the magnetic
pinch reported in G2006 to one SHARP resolution element ($\sim10$\arcsec
or $0.015\:\mathrm{pc}$). (2) deviations in the field appear as one
moves out beyond $20$\arcsec ($0.03\:\mathrm{pc}$) from 4A towards
4B. In the vicinity of 4B, the field morphology is rotated by $\sim30^{\circ}$
towards the horizontal with respect to the field orientation around
4A. (3) depolarization is observed towards the peak of 4B, as denoted
by open circles on the map where $p+2\sigma_{p}<1\%$. Here $p$ is
the degree of polarization and $\sigma_{p}$ is the corresponding
uncertainty in the value of $p$. Significant depolarization is not
observed towards 4A.

Taken together, these results suggest that 4A and 4B are magnetically
distinct objects. While at first glance the polarimetry of 4A appears
to be consistent with conventional ideas of magnetic support, 4B is
a more complicated case. The observed depolarization on 4B may result
from changes in the dust grain properties or shapes due to grain-growth
\citep{Vrba 1993, Hildebrand 1999}, supersonic and super-Alfv$\mathrm{\acute{e}}$nic
turbulence plus a lack of grain alignment above $A_{V}\approx3\:\mathrm{mag}$
\citep{Padoan 2001}, magnetic field geometry and the inclination
angle with the line-of-sight \citep{Goncalves 2005, Fiege 2000},
or finally because of beam smearing over small-scale field structures
\citep{Rao 1998}. The last point could be tested in a straightforward
manner with high-resolution polarimetry. In addition, \citet{Choi 2001}
identify a Class I object (denoted 4BII in the paper) within close
proximity to the younger 4B core. The existence of this more evolved
object is linked to the {}``Cav2'' cavity, located to the west of
the 4B complex, and may have generated this feature through the action
of an ancient outflow \citep{Choi 2001}. It is possible to speculate
that turbulence driven by this ancient outflow activity from 4BII
may have disrupted the embedded magnetic field and thus contributed
to the observed depolarization. We should note that no such Class
I objects have been identified within the 4A complex \citep{Choi 2005}. 

We now look at the position angles of the vectors situated within
$20$\arcsec~of the peak of 4A in order to assess the orientation
of the large-scale magnetic field around this source. We assume this
large-scale field has a simple, uniform morphology with a polarization
angle $\theta$. The value of $\theta$ can be calculated by computing
the mean of the individual orientation angles $\theta_{i}$, where
the subscript $i$ denotes the individual vectors. Table \ref{ta:pol}
contains all the relevant information on all the vectors depicted
in Figure 1. 

From the data presented in Table \ref{ta:pol}, a straight arithmetic
average of the orientation angles $\theta_{i}$ for the vectors situated
within a $20$\arcsec~radius of the peak of 4A yields a mean value
of $\theta\approx45.9^{\circ}$. The dispersion in the orientation
angles for these vectors ($\delta\theta_{\mathrm{obs}}$) is found
to be $\approx\pm13.6^{\circ}$. We therefore adopt $\theta\approx45.9^{\circ}$
as the orientation angle of the large-scale uniform magnetic field
around 4A. The G2006 result of $\approx61^{\circ}$ for their field
orientation is approximately a $1\sigma$ deviation from our result,
implying the two values are consistent with one another. Uncertainties
in our methods could account for the $\approx15^{\circ}$ discrepancy;
we admittedly assume a very simple uniform model for the field orientation,
ignoring any possible non-uniform large scale structure to the morphology.
The main outflow from 4A extends $\pm2$\arcmin with a position angle
of about $45^{\circ}$ for a line drawn from one tip of the outflow
to the other, but this rotates to $19^{\circ}$ for the inner part
of the outflow confined to a radius of approximately $40$\arcsec
\citep{Blake 1995, Choi 2005}. The reason for the change in orientation
of the outflow is unknown; cloud core rotation may have played a role.
A discussion of the full implications of this finding on the role
of rotational support in the formation of 4A is beyond the scope of
this paper. One would expect to see a significant deviation in the
field morphology from small to large spatial scales if centrifugal
forces were dominant in this system \citep{Machida 2006}. Instead,
the polarimetry results presented here plus those obtained at smaller
spatial scales (G2006) indicate otherwise. Nevertheless, the presence
of a binary system within 4A shows centrifugal forces could not be
negligible in the formation of this system. All of this information
is consistent with the idea that the magnetic and centrifugal forces
were comparable in magnitude for this system during the onset of collapse
(G2006). In addition, it should be noted that the large-scale uniform
magnetic field implied by our results is aligned with the original
(i.e., large-scale) orientation of the outflow from 4A.

Finally, we wish to calculate the mean intrinsic dispersion angle
($\delta\theta_{\mathrm{int}}$) over 4A by comparing our observations
with the adopted uniform magnetic field model with $\theta\approx45.9^{\circ}$.
Now $\delta\theta_{\mathrm{int}}$ is given by $\delta\theta_{\mathrm{int}}^{2}=\delta\theta_{\mathrm{obs}}^{2}-\sigma_{\theta}^{2}$,
where $\sigma_{\theta}$ is the uncertainty in the observed position
angle. We calculate values of $\delta\theta_{\mathrm{obs}}\approx13.6^{\circ}$
and $\sigma_{\theta}\approx7.7^{\circ}$ and as such work out $\delta\theta_{\mathrm{int}}$
to be $\approx11.2^{\circ}$. We will employ this value in our general
discussion in Section \ref{sec:3}. Note that we do not attempt a
similar analysis with our results over 4B. It is apparent from Figure
1 that we cannot fit our vectors over this source to a simple model
for the field orientation (plus no polarization is detected over the
peak flux of this source). As such, it is not possible to calculate
a meaningful average field orientation $\theta$ or intrinsic dispersion
$\delta\theta_{\mathrm{int}}$ for 4B.

\subsection{\label{sub:Spectroscopy}Spectroscopy}

Observations of the $\mathrm{HCN}$ $J=4\rightarrow3$ ($354.505\:\mathrm{GHz}$)
rotational transition from 4A were made with the CSO 300-400 GHz heterodyne
receiver. The beam size at this frequency is approximately $20$\arcsec~($0.029\:\mathrm{pc}$)
and thus samples a far larger region of space than was obtained with
DF2001. Detections were obtained for a sequence of points lying approximately
along the outflow axis of 4A, as well as for a single point that was
displaced from the center in the perpendicular direction. In total
seven different positions were looked at, the results are illustrated
in Figure 2.

What is immediately clear is the presence of a peak followed by a
dip in the line emission centered on 4A and the two positions lying
closest to it along the red-lobe of the outflow. We should note that
the presence of the red-lobe outflow will distort our spectra, as
the background that our source absorbs against is not flat but the
outflow emission itself. Therefore an interpretation of our data becomes
clearer once the outflow component of the spectra is fitted with a
Gaussian and removed, as was done using the CLASS software package
and is shown in Figure 3. The spectra in Figure 3 are characterised
solely by the aforementioned emission peak and dip, with the dip being
situated to the right of the peak in each case. Each maxima and minima
are observed at velocities of $6.8\:\mathrm{km/s}$ and $8.0\:\mathrm{km/s}$,
respectively. The full-width-half-maximum of each peak and dip profile
is approximately $\simeq1.5\:\mathrm{km/s}$, and the point between
the maxima and minima where the line temperature $T_{B}$ equals $\approx0\:\mathrm{K}$
corresponds to a velocity of approximately $V\approx7\:\mathrm{km/s}$
in each case. These data have exactly the same characteristics as
the $\mathrm{H_{2}CO}\:3_{12}\rightarrow2_{11}$ spectra presented
in Figure 4 of DF2001 and are thus characteristic of inverse P-Cygni
profiles. With this interpretation we note that the spectral signatures
seen in Figure 2 are characteristic of an infalling envelope of gas
around 4A plus a outflow. These features are not observed towards
the blue-lobe of the outflow. This is to be expected due to the position
of the blue-lobe outflow, which is situated in between the infalling
envelope and the observer. Therefore, this component of the outflow
does not provide a background against which the infalling material
can absorb. We do not have spectroscopy data on 4B at this time, and
as such we cannot comment on the nature of gas motions around this
core.

One also notices the disappearance of the inverse P-Cygni profile
at a offset position of $\alpha_{o}=27$\arcsec, $\delta_{o}=40$\arcsec~along
the red-lobe outflow. We can therefore state that the peak of 4A is
surrounded by an infalling envelope of radius $r_{c}$ on the plane
of the sky, where $0.05\:\mathrm{pc}\:\leq\: r_{c}\:<\:0.07\:\mathrm{pc}$
. The lower bound of $r_{c}$ is provided by the spectra obtained
at an offset position of $\alpha_{o}=15$\arcsec, $\delta_{o}=30$\arcsec;
the furthest position away from the peak of 4A at which the infall
signature is still clearly apparent. The range of $r_{c}$ specified
here is consistent with the value of $\approx0.1\:\mathrm{pc}$ predicted
by models of self-gravitating cores \citep{McKee 2007}. Note that
this estimate relies upon the assumption that nothing obstructs our
view of the infall signature at a radial distance $\approx0.07\:\mathrm{pc}$
from the peak flux of 4A. Adopting $r_{c}\approx0.06\:\mathrm{pc}$,
we immediately note that the size of the infalling envelope is approximately
$\sim4$ times larger than the upper limit placed on the size of the
magnetic pinch observed around 4A (see Section \ref{sub:Polarimetry}).
This result is consistent with theoretical work on the collapse of
magnetized cloud cores \citep{Galli 1993}.

\section{\label{sec:3}Discussion}

\subsection{Kinematics}

To further our investigation of the 4A system we fit our three inverse
P-Cygni profile detections to an enhanced version of the \textquotedblleft{}two-layer\textquotedblright{}
model originally devised by \citet{Myers 1996} and later used by
DF2001. In its most general form, our treatment of this problem is
to envisage two parallel layers of material moving towards an opaque
central condensation. These two layers, denoted {}``front slab''
and {}``rear slab'', represent the infalling envelope. The opaque
source is taken to fill a fraction $\Upsilon$ of the telescope beam.
This setup is the model of DF2001, where the presence of outflows
was not accounted for. For our model, two more layers of material
are included in this system to represent the blue- and red-lobe outflows.
This scenario is illustrated in Figure 4. 

Quantifying this model into an expression of the line brightness temperature
$T_{\mathrm{B}}\left(V\right)$ at a certain velocity $V$ we obtain
equation (1) below. Note the subscripts \textquoteleft{}$B$\textquoteright{},
\textquoteleft{}$R$\textquoteright{}, \textquoteleft{}$f$\textquoteright{},
\textquoteleft{}$r$\textquoteright{}, \textquoteleft{}$c$\textquoteright{},
and \textquoteleft{}$bg$\textquoteright{} represent the blue-lobe
outflow, red-lobe outflow, front slab, rear slab, central source,
and cosmic background parameters, respectively. Therefore the analytical
model for the scenario depicted in Figure 4 is:

\begin{equation}
\begin{array}{ccc}
T_{\mathrm{B}}\left(V\right) & = & \left(1-\Upsilon\right)\cdot J\left(T_{R}\right)\cdot\left(1-e^{-\tau_{R}}\right)\cdot e^{-\tau_{0}}+\left(1-\Upsilon\right)\cdot J\left(T_{r}\right)\cdot\left(1-e^{-\tau_{r}}\right)\cdot e^{-\tau_{f}-\tau_{B}}\\
 &  & +J\left(T_{f}\right)\cdot\left(1-e^{-\tau_{f}}\right)\cdot e^{-\tau_{B}}+J\left(T_{B}\right)\cdot\left(1-e^{-\tau_{B}}\right)\\
 &  & -\Upsilon\cdot J\left(T_{c}\right)\cdot\left(1-e^{-\tau_{f}-\tau_{B}}\right)-\left(1-\Upsilon\right)\cdot J\left(T_{bg}\right)\cdot\left(1-e^{-\tau_{\star}}\right),\end{array}\end{equation}
where $J\left(T\right)=T_{0}/\left(e^{T_{0}/T}-1\right)$ is the Planck
temperature as a function of the blackbody temperature $T$, and $T_{0}=h\upsilon/k$
where $h$ is the Planck constant, $k$ is the Boltzmann constant,
and $\upsilon$ is the frequency. The optical depth for model component
$i$ is denoted as $\tau_{i}$. Finally we note that $\tau_{0}=\tau_{r}+\tau_{f}+\tau_{B}$
and $\tau_{\star}=\tau_{R}+\tau_{0}$. Following \citet{Myers 1996}
we model the different optical depths with Gaussian profiles:

\begin{equation}
\tau_{i}=\tau_{0i}\cdot exp\left[\frac{-\left(V-V_{i}-V_{\mathrm{LSR}}\right)^{2}}{2\sigma_{i}^{2}}\right]\end{equation}

\begin{equation}
\tau_{j}=\tau_{0j}\cdot exp\left[\frac{-\left(V+V_{j}-V_{\mathrm{LSR}}\right)^{2}}{2\sigma_{j}^{2}}\right],\end{equation}
where $i=f,\, R$ and $j=r,\, B$, $\tau_{0i}$ and $\tau_{0j}$ are
the peak optical depths, $\sigma_{i}$ and $\sigma_{j}$ are the respective
velocity dispersions, and the velocity for the local standard of rest
is taken to be $V_{\mathrm{LSR}}\approx6.96\:\mathrm{km/s}$ (DF2001).
Note that $V_{f}$ and $V_{r}$ are the infall velocities for the
front and rear slabs, respectively. 

We can simplify (1) by making note of two important properties of
our data. First, our large beam size implies $\Upsilon\approx0$ (DF2001
get a value of $\Upsilon\approx0.3$ with a $2$\arcsec~beam). Second,
we neglect the contribution of the blue-lobe, as only the red-lobe
component of the outflow provides the background radiation against
which the infalling material can absorb. It is clear from Figure 2
that the spectra centered on ($\alpha_{o}=0$\arcsec, $\delta_{o}=0$\arcsec)
has a significant blue-lobe outflow component while the spectra centered
on ($\alpha_{o}=7$\arcsec, $\delta_{o}=20$\arcsec) and ($\alpha_{o}=15$\arcsec,
$\delta_{o}=30$\arcsec) are largely dominated by the red-lobe outflow.
As such, our outflow approximation will be especially coarse in the
case of the spectra centered on the peak flux of 4A. By setting $\Upsilon=0$
and $\tau_{B}=0$ in equation (1) we get: 

\begin{equation}
\begin{array}{ccc}
T_{\mathrm{B}}\left(V,\tau_{B}=0\right) & = & J\left(T_{R}\right)\cdot\left(1-e^{-\tau_{R}}\right)\cdot e^{-\tau_{f}-\tau_{r}}+J\left(T_{r}\right)\cdot\left(1-e^{-\tau_{r}}\right)\cdot e^{-\tau_{f}}\\
 &  & +J\left(T_{f}\right)\cdot\left(1-e^{-\tau_{f}}\right)-J\left(T_{bg}\right)\cdot\left(1-e^{-\tau_{\oplus}}\right),\end{array}\end{equation}
where $\tau_{\oplus}=\tau_{f}+\tau_{r}+\tau_{R}$. A computer program
was written to minimize the reduced-$\chi^{2}$ function generated
by comparing equation (4) with a multi-Gaussian fit to the three aforementioned
spectra in Figure 2 that exhibit inverse P-Cygni profiles. The minimization
was carried out through the use of Powell's Method for multidimensional
functions \citep{nr}. The resulting fit of equation (4) to our spectra
at offset positions ($\alpha_{o}=0$\arcsec, $\delta_{o}=0$\arcsec),
($\alpha_{o}=7$\arcsec, $\delta_{o}=20$\arcsec), and ($\alpha_{o}=15$\arcsec,
$\delta_{o}=30$\arcsec) are shown in Figure 5. 

The plots shown in Figure 5 demonstrate a reasonably good agreement
between the actual data and the model. We should take stock at this
point to stress that this model provides the simplest possible explanation
for a contracting system with outflow. As such, it will provide us
with only an approximate picture of the physical properties at play
in 4A and its surroundings. This is especially true with regards to
the optical depths and temperatures of the absorbing/emitting gas
(DF2001). It should also be mentioned that the resolution of our data
prevents us from distinguishing between small-scale structures, such
as the binary system at the core of the 4A complex (dubbed 4A1 and
4A2, G2006). Although we dispense with the optically thick central
condensation in equation (4), each member of the binary system drives
a unique outflow \citep{Choi 2005}%
\footnote{Note that all referrals to the {}``4A outflow'' made elsewhere in
this paper pertain to the one driven by 4A2.%
}. By far the dominant outflow originates from 4A2, which possesses
a length $\sim11$ times longer and is more luminous than the outflow
associated with 4A1 \citep{Choi 2005, Blake 1995}. Our spectra likely
sample an average of the environment of 4A, with detections of both
outflows being integrated for data obtained on the peak flux of 4A.
However, it is unlikely the data obtained away from the peak will
be affected by the outflow of 4A1, since this outflow is limited in
extent and has a position angle shifted by $\sim20^{\circ}$ to the
position angle of the much larger 4A2 outflow. Despite these shortcomings,
our model is still useful for the comparison of the infall velocity
between different spectra, since the ratio $V_{f}/\sigma_{f}$ (or
$V_{r}/\sigma_{r}$) is a key parameter for any model of a contracting
system (DF2001; Leung \& Brown 1977). 

The model parameters resulting from the simulations shown in Figure
5 are listed in Table \ref{ta:spe}. One notices immediately that
$V_{f}>V_{r}$ for each scan, where the values of $V_{f}$ tend to
be close to a speed of $\sim1\:\mathrm{km/s}$ while $V_{r}$ tends
to have values closer to $\sim0.3\:\mathrm{km/s}$. A likely explanation
for this is that the value of $V_{\mathrm{LSR}}$ set in our model
is somewhat inaccurate, and thus introduces a systematic error in
our calculations. We note DF2001 leave $V_{\mathrm{LSR}}$ as a free
parameter in their version of the {}``two-layered'' model. Our attempt
to treat $V_{\mathrm{LSR}}$ as a free parameter in equation (4) failed
to produce satisfactory results. This may be due to the highly non-linear
nature of our model, which makes an iterative fit of equation (4)
to a particular data set very sensitive to the initial parameter values.
The simplest means to proceed was by fixing $V_{\mathrm{LSR}}$ to
the DF2001 value. The differences between equation (4) and the DF2001
model may account for the different $V_{\mathrm{LSR}}$ value in our
case. To correct for this we take an arithmetic average of the $V_{f}$
and $V_{r}$ values listed in Table \ref{ta:spe}, which result in
mean infall velocities of $0.63\:\mathrm{km/s}$, $0.61\:\mathrm{km/s}$,
and $0.67\:\mathrm{km/s}$ for the scans at offset positions ($\alpha_{o}=0$\arcsec,
$\delta_{o}=0$\arcsec) , ($\alpha_{o}=7$\arcsec, $\delta_{o}=20$\arcsec),
and ($\alpha_{o}=15$\arcsec, $\delta_{o}=30$\arcsec), respectively.
Taking all three scans together we estimate a mean infall velocity
of $\approx0.64\:\mathrm{km/s}$ for 4A, a value that is very similar
to the $0.68\:\mathrm{km/s}$ velocity calculated by DF2001. If we
assume a gas temperature of $T_{g}\approx30\:\mathrm{K}$ \citep{Blake 1995}
we can calculate the isothermal sound speed $V_{\mathrm{rms}}=\sqrt{kT_{g}/\mu m_{\mathrm{H}}}$
to be $\approx0.33\:\mathrm{km/s}$, where $k$ is Boltzmann's constant,
$\mu$ is the mean molecular weight ($\mu=2.22$), and $m_{\mathrm{H}}$
is the mass of hydrogen. Thus the observed infall is also supersonic. 

Assuming inside-out collapse for 4A, we note that the radius $r_{c}$
of the expansion wave for the infalling envelope is given by $r_{c}=V_{\mathrm{rms}}\times t$,
where $V_{\mathrm{rms}}$ is defined above and $t$ is the time since
the onset of collapse. Taking $r_{c}\approx0.06\:\mathrm{pc}$ (see
Section \ref{sub:Spectroscopy}), we find $t\approx2\times10^{5}\:\mathrm{yr}$.
A large degree of uncertainty exists with regards to the age of the
outflow, but estimates range from $2000-20000\:\mathrm{yr}$ \citep{Blake 1995}.
Therefore the age of the infalling envelop in 4A is roughly $10-100$
times the age of the associated outflow. 

Outflow velocities along the line-of-sight of $-2.86\mathrm{\: km/s}$,
$10.21\mathrm{\: km/s}$, and $9.17\mathrm{\: km/s}$ are estimated
from the fits of equation (4) to the scans at offset positions ($\alpha_{o}=0$\arcsec,
$\delta_{o}=0$\arcsec) , ($\alpha_{o}=7$\arcsec, $\delta_{o}=20$\arcsec),
and ($\alpha_{o}=15$\arcsec, $\delta_{o}=30$\arcsec), respectively.
The latter two values indicate the red-lobe outflow has an approximate
velocity of $\approx10\:\mathrm{km/s}$. The negative value for the
velocity at position ($\alpha_{o}=0$\arcsec, $\delta_{o}=0$\arcsec)
is to due to the fact that equation (4) treats for a red-lobe outflow
only, while the spectrum at this position shows a significant blue-lobe
outflow as well. The velocity values for both red-lobe outflows are
approximately consistent with $\mathrm{SiO}$ $J=1\rightarrow0$ observations
presented in \citet{Choi 2005} and illustrated in Figure 5 of that
paper. The outflow observed at position ($\alpha_{o}=0$\arcsec,
$\delta_{o}=0$\arcsec) is a composite of both the red- and blue-lobes
and hence our model does not provide a reliable outflow velocity for
this case.

As a final note we wish to briefly discuss our handling of the outflow
velocity $V_{R}$ employed in our model for the scan at position ($\alpha_{o}=0$\arcsec,
$\delta_{o}=0$\arcsec). Because of the simplifications introduced
into equation (4), where only the red-lobe outflow is included analytically,
it is necessary to set this parameter to a negative value in order
to fit the composite red- and blue-lobe outflow that are present in
this spectrum. Although this points out a shortcoming of our model,
the important point here is to maintain emitting gas in the background
of the infalling envelope; a red-lobe outflow. This component of the
model provides emission from a backdrop of material against which
the infalling gas can absorb and thus produce the dip present in a
inverse P-Cygni profile. This is the reason why $V_{R}$ is negative
in equation (4) at position ($\alpha_{o}=0$\arcsec, $\delta_{o}=0$\arcsec).
Despite this obvious defect, the model used here is the simplest mathematical
construct that still conveys physical meaning for the data at hand.
Future work may wish to employ more sophisticated Monte Carlo techniques
to generate models that more accurately describe this source.

\subsection{Support Mechanism}

We now proceed to calculate a mass estimate for 4A and 4B. This can
be achieved through a careful treatment of the thermal dust continuum
data from SHARP. The mass of a cloud can be estimated via its submillimeter
thermal emission by: 

\begin{equation}
M=\frac{4}{3}\varrho d^{2}\frac{F_{\nu}}{B_{\nu}\left(T_{d}\right)}\frac{a}{\varepsilon_{\nu}}R,\end{equation}
where $R$ is the gas to dust ratio, $\varrho$ is the density of
the dust material, $d$ is the distance to the cloud, $F_{\nu}$ is
the measured flux density, $B_{\nu}\left(T_{d}\right)$ is the Planck
function with dust temperature $T_{d}$, $a$ is the grain radius,
and $\varepsilon_{\nu}$ is the dust emissivity \citep{Hildebrand 1983}.
Here we assume values of $R=100$, $\varrho=3\:\mathrm{g/cm^{3}}$,
$d=300\:\mathrm{pc}$, $a=10^{-5}\:\mathrm{cm}$, $\varepsilon_{\nu}=2.26\times10^{-4}$,
and $T_{d}=50\:\mathrm{K}$, where the dust temperature was chosen
to be the same value as that employed in G2006. The values of $\varrho$,
$a$, and $\varepsilon_{\nu}$ were taken or inferred from \citet{Hildebrand 1983}.
The total amount of $350\:\mathrm{\mu m}$ flux detected within a
radial distance of $20$\arcsec ($0.03\:\mathrm{pc}$) from 4A is
$\approx179\:\mathrm{Jy}$, while a total of $\approx76\:\mathrm{Jy}$
was detected within $10$\arcsec ($0.015\:\mathrm{pc}$) of 4B. The
distance around 4A was selected to encompass the polarimetry vectors
used in the calculation of $\delta\theta_{\mathrm{int}}$ in Section
\ref{sub:Polarimetry}. With this information, the mass of 4A and
4B is estimated to be $\approx1.9\:\mathrm{M_{\odot}}$ and $\approx0.8\:\mathrm{M_{\odot}}$,
respectively. These values are comparable to the mass estimates of
DF2001 and G2006 that are stated in Section \ref{sec:Introduction}
of this work. As a final point we note that the total flux detected
in Figure 1 is $\approx511\:\mathrm{Jy}$. From this flux value a
total detected mass of $\approx5.4\:\mathrm{M_{\odot}}$ is calculated
for the area of NGC 1333 IRAS 4 mapped by our continuum observations. 

We can now apply a modified form of the Chandrasekhar \& Fermi (CF)
technique \citep{CF 1953} to obtain an estimate of the magnetic field
strength around 4A. If we assume the dispersion in our magnetic field
map (see Figure 1B) is entirely due to Alfv$\mathrm{\acute{e}}$n
waves and/or turbulence, then the plane-of-sky magnetic field strength
will be given by:

\begin{equation}
\begin{array}{ccc}
B_{\mathrm{pos}} & = & Q\left(\frac{\delta v_{\mathrm{los}}}{\delta\theta_{\mathrm{int}}}\right)\left(4\pi\rho\right)^{1/2}.\end{array}\end{equation}
Here we take $Q=0.5$, the same value used in G2006 and one that is
indicated by simulations of turbulent molecular clouds \citep{Ostriker 2001}.
We have already calculated $\delta\theta_{\mathrm{int}}\approx11.2^{\circ}$
in Section \ref{sec:2}. The gas density $\rho$ can be roughly estimated
by dividing the mass of 4A calculated above by the volume of a $20$\arcsec
($0.03\:\mathrm{pc}$) radius sphere. Doing this yields a density
of $\rho\approx8\times10^{-19}\:\mathrm{g/cm^{3}}$. The line-of-sight
velocity dispersion $\delta v_{\mathrm{los}}$ is taken from observations
of $\mathrm{HCO^{+}}$ $J=4\rightarrow3$ obtained at the same time
(and using the same heterodyne receiver) as the $\mathrm{HCN}$ observations
discussed earlier. These data were obtained while pointing on the
peak flux of 4A and are shown in Figure 6. The choice of analyzing
this spectrum for the $\delta v_{\mathrm{los}}$ value is motivated
by the fact that an ion will be better coupled to the magnetic field
(and the dust) than a neutral species over the whole turbulent energy
density spectrum \citep{Li and Houde 2008}. Thus the line width of
this species will be more representative of the turbulence/Alfv$\acute{\mathrm{e}}$n
waves that may be disturbing the field as opposed to a neutral molecular
counterpart. The value of $\delta v_{\mathrm{los}}$ is found to be
$1.67\:\mathrm{km/s}$. Inserting this information into equation (6)
yields a field strength of $B_{\mathrm{pos}}\approx1.4\:\mathrm{mG}$,
a value that is roughly a factor of $3.6$ lower than the G2006 result.
This difference is not unexpected since the field strength should
increase towards the center of the core, as this is the location of
maximum compression of the field lines. One could thus expect the
value of $B_{\mathrm{pos}}$ to diminish at larger spatial scales. 

We are now in a position to calculate the mass-to-flux ratio for our
observations on 4A. Writing this quantity in terms of the critical
value for collapse $\lambda$ we find:

\begin{equation}
\begin{array}{ccc}
\lambda & = & \frac{M/\Phi}{c_{\Phi}/\sqrt{G}}\end{array}\end{equation}
where $c_{\Phi}=0.12$ \citep{Tomisaka 1988}, $\Phi$ is the magnetic
flux, and $G$ is the gravitational constant. Using this relation
we calculate $\lambda\approx0.44$ over a region of $\approx270\:\mathrm{arcsec^{2}}$
centered on the peak of 4A. This value is a factor of $\approx2.3$
lower than unity and also notably lower than the G2006 result. A strongly
sub-critical cloud is also inconsistent with our own observations
of infalling gas motions (Section \ref{sub:Spectroscopy}). Uncertainties
in our estimate of $M$ may be partly responsible; submillimeter continuum
emission will fail to sample the hot protostellar mass where dust
particles cannot exist. We should note that the use of the plane-of-sky
component of the magnetic field will result in an underestimate of
$\Phi$ and thus lead to a overestimation of $\lambda$. Recent work
suggests the inclination angle $\vartheta$ of the embedded magnetic
field in this cloud falls within the range $0^{\circ}\leq\vartheta\leq55^{\circ}$
\citep{Goncalves 2008}. This suggests that the magnitude of $B=B_{\mathrm{pos}}/cos\vartheta$
could fall somewhere within $1.4\:\mathrm{mG}\leq B\leq2.4\:\mathrm{mG}$.
Therefore we can speculate that this effect could result in an overestimation
of $\lambda$ by up to a factor of $\approx1.7$.

A more likely explanation for the sub-critical value of $\lambda$
may come from the application of the CF technique itself. Crucial
to this calculation is an accurate measure of the intrinsic dispersion
angle $\delta\theta_{\mathrm{int}}$ of the magnetic field vectors.
Polarimetry, however, does not distinguish between contributions along
the line-of-sight; the result being that the angular dispersion will
be reduced through the process of signal integration through the thickness
of the cloud as well as across the area subtended by the telescope
beam \citep{Hildebrand 2009}. The value of $\delta\theta_{\mathrm{int}}$
that is then calculated could be smaller than the true dispersion
and thus may result in an overestimation of $B_{\mathrm{pos}}$ through
equation (6). Despite the fact that the factor $Q$ in equation (6)
is meant to account for the problem of signal integration through
the cloud \citep{Ostriker 2001}, in reality this is only a first-order
correction and the smoothing effect could be more severe \citep{Houde 2009}.
Higher resolution polarimetry across the large spatial scales observed
here in conjunction with a more in-depth analysis may be used in the
future to resolve this issue \citep{Hildebrand 2009, Houde 2009}. 

We can now calculate the ratio of the turbulent to magnetic energy
($\beta_{\mathrm{turb}}$) in 4A. The value we determine for this
quantity is $\beta_{\mathrm{turb}}\approx8\times10^{-4}\left(\delta\theta_{\mathrm{int}}/1^{\circ}\right)^{2}=0.05$,
suggesting magnetic forces dominate the physics of this particular
cloud. From our discussion above we note that the field strength calculated
here could be overestimated by a factor of a few. This could result
from an underestimate of the intrinsic dispersion in the polarimetry
$\delta\theta_{\mathrm{int}}$. For example, a requirement for a critical
value of $\lambda$ would be an intrinsic dispersion value of $\delta\theta_{\mathrm{int}}\approx26^{\circ}$.
This would then correspond to a value of $\beta_{\mathrm{turb}}\approx0.5$,
a value sufficiently close to unity to suggest an equipartition of
the turbulent and magnetic energies in this system. These results
are definitely not consistent with is the idea of a turbulent-dominated
support system for this cloud. Since models of magnetically-supported
star formation predict $\lambda\approx1$ for cloud cores undergoing
collapse, we conclude that our results are likely consistent with
the idea that the turbulent and magnetic energies are of the same
order of magnitude in this system.

\section{\label{sec:4}Conclusions}

We summarize our results as follows:
\begin{itemize}
\item Dust continuum polarimetry done at $350\:\mathrm{\mu m}$ with SHARP
demonstrates a uniform magnetic field morphology around 4A at a resolution
scale of $10$\arcsec ($0.015\:\mathrm{pc}$). We therefore adopt
the size of one resolution element to be an approximate upper limit
on the size of the magnetic pinch region reported in G2006. This is
in agreement with magnetic cloud collapse theory and correlates well
with the findings of G2006. In addition, the large-scale uniform magnetic
field appears to be aligned with the original (large-scale) outflow
direction for 4A. Why the large-scale and small-scale directions of
the 4A outflow are different is not known, but our results and those
of G2006 show that the average orientation of the magnetic field has
not changed direction from large-scales to small. The polarimetry
obtained on 4B appears to indicate depolarization towards the peak
flux region. An explanation for these observations on 4B remains an
open point. Mass estimates for both 4A and 4B have been made revealing
$1.9\:\mathrm{M_{\odot}}$ within $20$\arcsec ($0.03\:\mathrm{pc}$)
and $0.8\:\mathrm{M_{\odot}}$ within $10$\arcsec ($0.015\:\mathrm{pc}$)
of the peak flux for both cores, respectively. The total mass traced
by our continuum observations is $5.4\:\mathrm{M_{\odot}}$ for the
portion of the NGC 1333 IRAS 4 cloud complex surveyed. 
\item Spectroscopy done at the CSO with $\mathrm{HCN}$ $J=4\rightarrow3$
line emission has revealed inverse P-Cygni profiles at offset positions
($\alpha_{o}=0$\arcsec, $\delta_{o}=0$\arcsec), ($\alpha_{o}=7$\arcsec,
$\delta_{o}=20$\arcsec), and ($\alpha_{o}=15$\arcsec, $\delta_{o}=30$\arcsec)
from the location of peak flux for 4A. Fitting these data with an
enhanced version of the \citet{Myers 1996} \textquotedblleft{}two-layer\textquotedblright{}
model, we estimate a mean infall speed of $0.64\:\mathrm{km/s}$ for
this cloud core. These findings are in good agreement with the results
of DF2001. The radial size of the infalling envelope $r_{c}$ is estimated
to range between $0.05\:\mathrm{pc}\:\leq\: r_{c}\:<\:0.07\:\mathrm{pc}$
and is thus approximately $\sim4$ times larger than the size of the
magnetic pinch. This is consistent with theoretical work on magnetized
cloud collapse \citep{Galli 1993}. The age of the infalling envelope
is found to be approximately $\approx2\times10^{5}\:\mathrm{yr}$;
a figure that is roughly $10-100$ times larger than the age of the
associated outflow.
\item The value of the plane-of-sky component of the magnetic field strength
has been estimated to be $\approx1.4\:\mathrm{mG}$ around 4A. This
yields a normalized mass to flux ratio of $\lambda\approx0.44$. This
value of $\lambda$ is inconsistent with our observations of infalling
gas motions and does not agree with the previous results of G2006.
We speculate here that $\delta\theta_{\mathrm{int}}$ may be underestimated
due to smoothing effects in the angular dispersion along the line-of-sight
that is characteristic of polarimetry. If we inflate our value of
$\delta\theta_{\mathrm{int}}$ by a factor of $1/\lambda$ , we then
find that $\beta_{\mathrm{turb}}\approx0.5$. This suggests an equipartition
of the magnetic and turbulent energies in this cloud. 
\end{itemize}

\acknowledgements{}

Our group is grateful for the assistance of the Caltech Submillimeter
Observatory staff in installing and observing with SHARP and the heterodyne
receiver. We also acknowledge the help provided by Jose Girart, Megan
Krejny, Roger Hildebrand, Tristan Matthews, Larry Kirby, and Lerothodi
Leeuw. M.A.'s and M.H.'s research is funded through the NSERC Discovery
Grant, Canada Research Chair, Canada Foundation for Innovation, Ontario
Innovation Trust, and Western's Academic Development Fund programs.
SHARC II is funded through the NSF grant AST 05-40882 to the California
Institute of Technology. The development of SHARP was funded by an
NSF grant to Northwestern University (AST 02-43156), and its subsequent
commissioning was funded by NSF grants to Northwestern University
(AST 05-05230) and University of Chicago (AST 05-05124).

\clearpage

\begin{figure}
\epsscale{0.5}
\rotate
\plotone{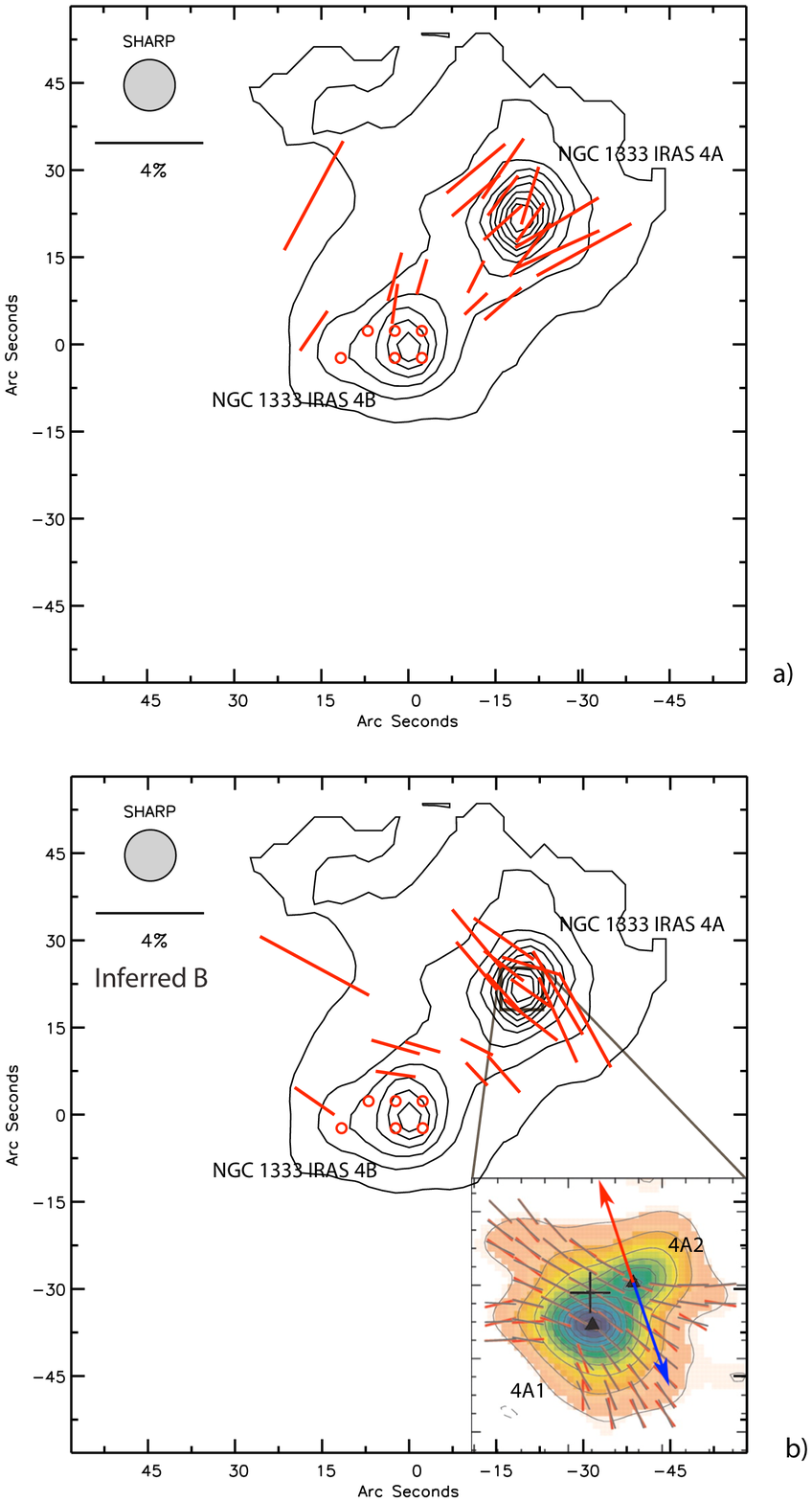}

\caption{SHARP polarimetry (a) and deduced magnetic field orientation (b) over
4A and 4B. Both images are centered on 4B ($\alpha_{2000}=3h29m12.06s$,
$\delta_{2000}=+31^{\circ}13$\arcmin$10.8$\arcsec) with 4A lying
towards the northwest corner of the map. The horizontal and vertical
axes show offsets in Right Ascension and Declination, respectively.
Contour levels are 0.1, 0.2, \ldots{} 0.9 times the peak flux value
($29.3\:\mathrm{Jy/beam}$). Image b) also shows the G2006 magnetic
field map for comparison, where 4A is resolved into its components
4A1 and 4A2. Arrows indicate the orientation of the outflow. Note
that all the vectors presented in image a) are $p>3\sigma_{p}$ and
circles denote regions where $p+2\sigma_{p}<1\%$.}

\end{figure}

\begin{figure}
\epsscale{0.5}
\plotone{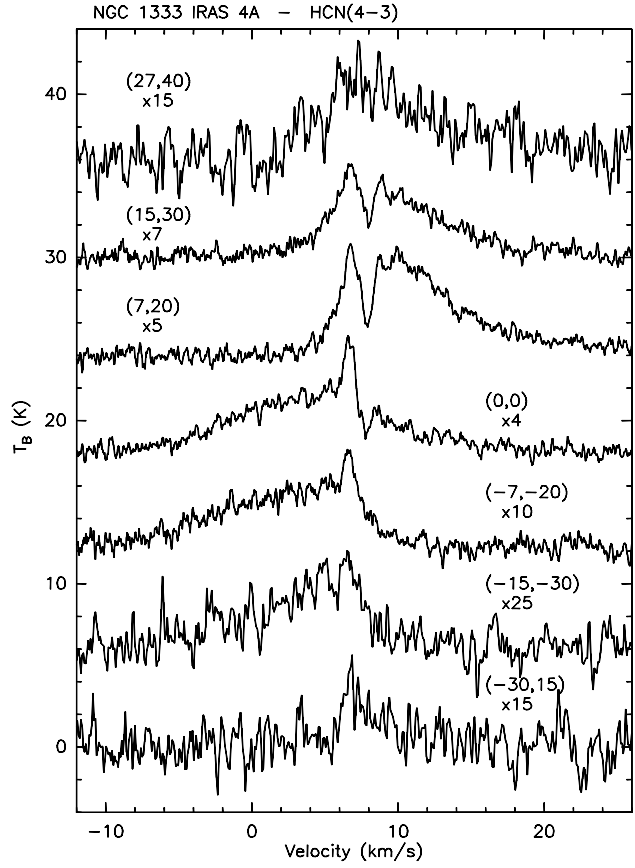}

\caption{$\mathrm{HCN}\: J=4\rightarrow3$ line emission ($354.505\:\mathrm{GHz}$)
centered on 4A ($\alpha_{2000}=3h29m10.42s$, $\delta_{2000}=+31^{\circ}13$\arcmin$35.4$\arcsec).
Observations were made out to offset positions of ($\alpha_{o}=-15$\arcsec,
$\delta_{o}=-30$\arcsec) on the blue lobe of the outflow and ($\alpha_{o}=27$\arcsec,
$\delta_{o}=40$\arcsec) on the red lobe. One observation at an offset
position of ($\alpha_{o}=-30$\arcsec, $\delta_{o}=15$\arcsec)
was made perpendicular to the outflow axis.}

\end{figure}

\begin{figure}
\epsscale{0.7}
\plotone{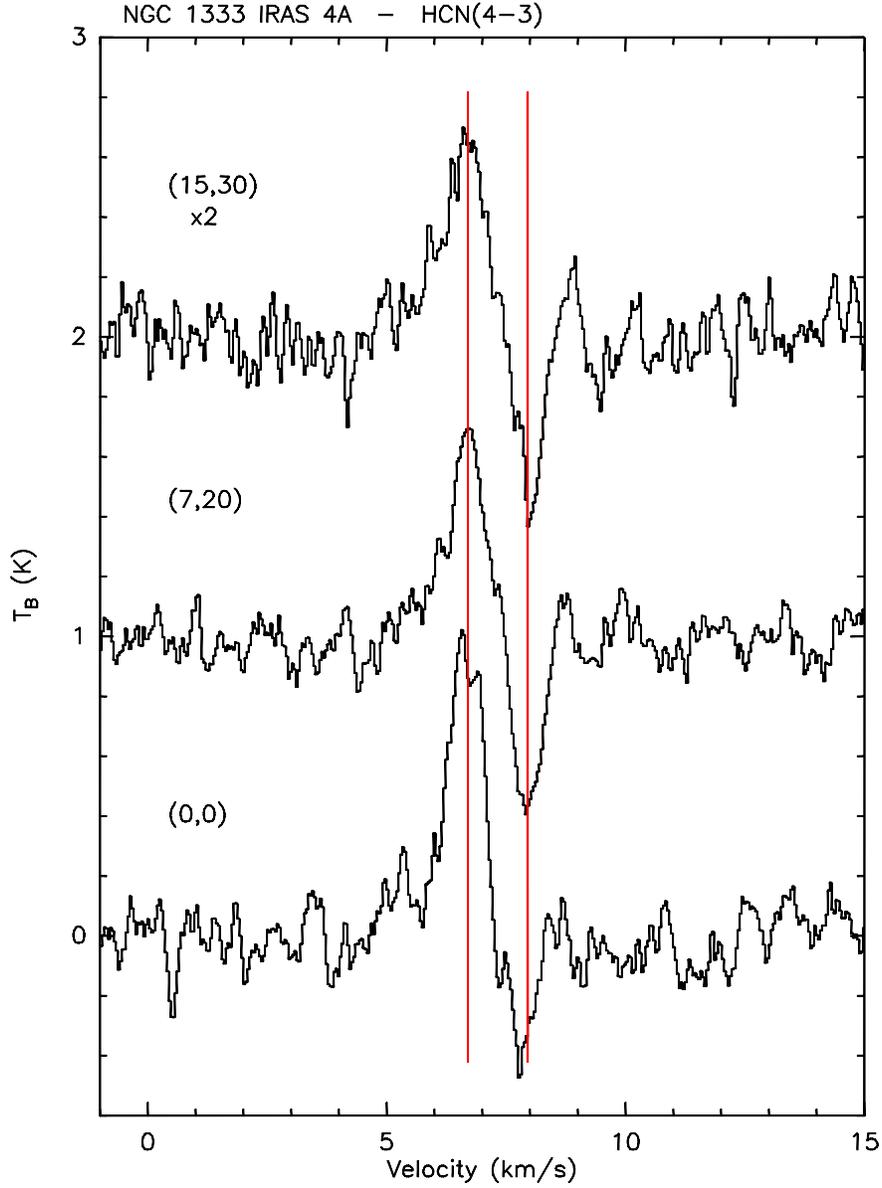}

\caption{$\mathrm{HCN}\: J=4\rightarrow3$ line emission at offset positions
($\alpha_{o}=0$\arcsec, $\delta_{o}=0$\arcsec), ($\alpha_{o}=7$\arcsec,
$\delta_{o}=20$\arcsec), and ($\alpha_{o}=15$\arcsec, $\delta_{o}=30$\arcsec)
from the peak flux of 4A. Outflow components have been removed to
reveal characteristic inverse P-Cygni profiles at each of the three
locations. The outflows were removed with the use of the CLASS software
package. Vertical lines represent velocities at $6.8\:\mathrm{km/s}$
and $8.0\:\mathrm{km/s}$.}

\end{figure}

\begin{figure}
\epsscale{0.5}
\plotone{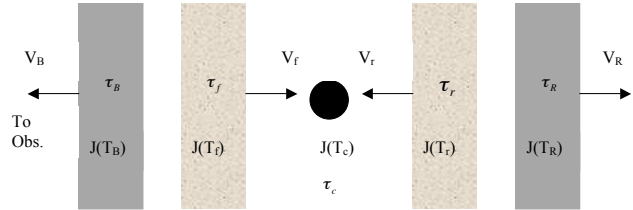}

\caption{Illustration of the enhanced \textquotedblleft{}four-layered\textquotedblright{}
version of the two-layered model of \citep{Myers 1996}. This depiction
labels each component of the model with a Planck temperature $J\left(T_{i}\right)$,
optical depth $\tau_{i}$, and a velocity $V_{i}$. The subscript
$i$ denotes \textquoteleft{}$B$\textquoteright{}, \textquoteleft{}$R$\textquoteright{},
\textquoteleft{}$f$\textquoteright{}, \textquoteleft{}$r$\textquoteright{},
and \textquoteleft{}$c$\textquoteright{} representing the blue outflow,
red outflow, front slab, rear slab, and central source parameters
respectively. Arrows indicate velocity directions along the observers
line-of-sight. We also allow for cosmic background radiation with
a Planck temperature $J\left(T_{bg}\right)=0.49\:\mathrm{K}$.}

\end{figure}

\begin{figure}
\epsscale{0.5}
\plotone{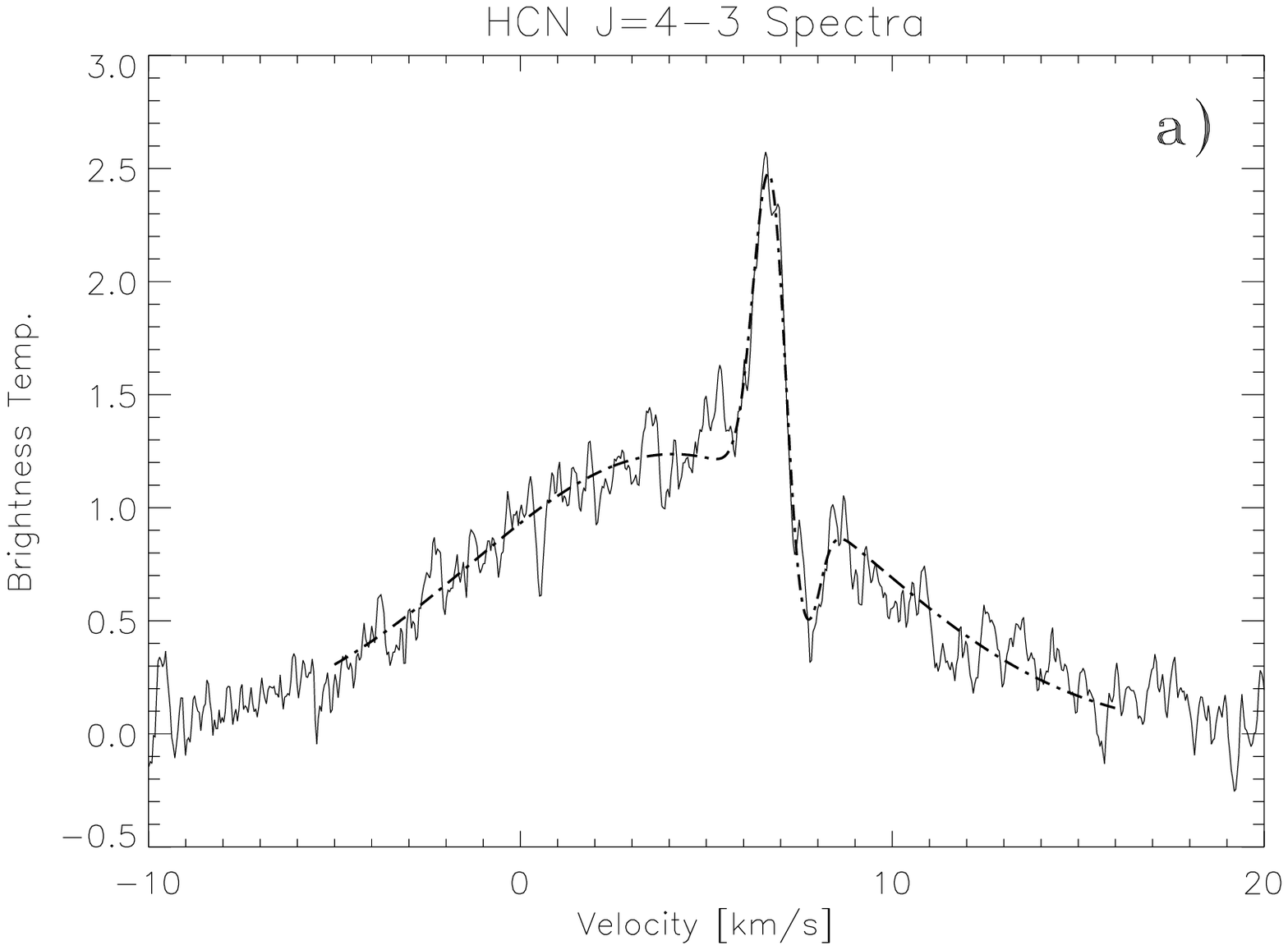}

\epsscale{0.5}
\plotone{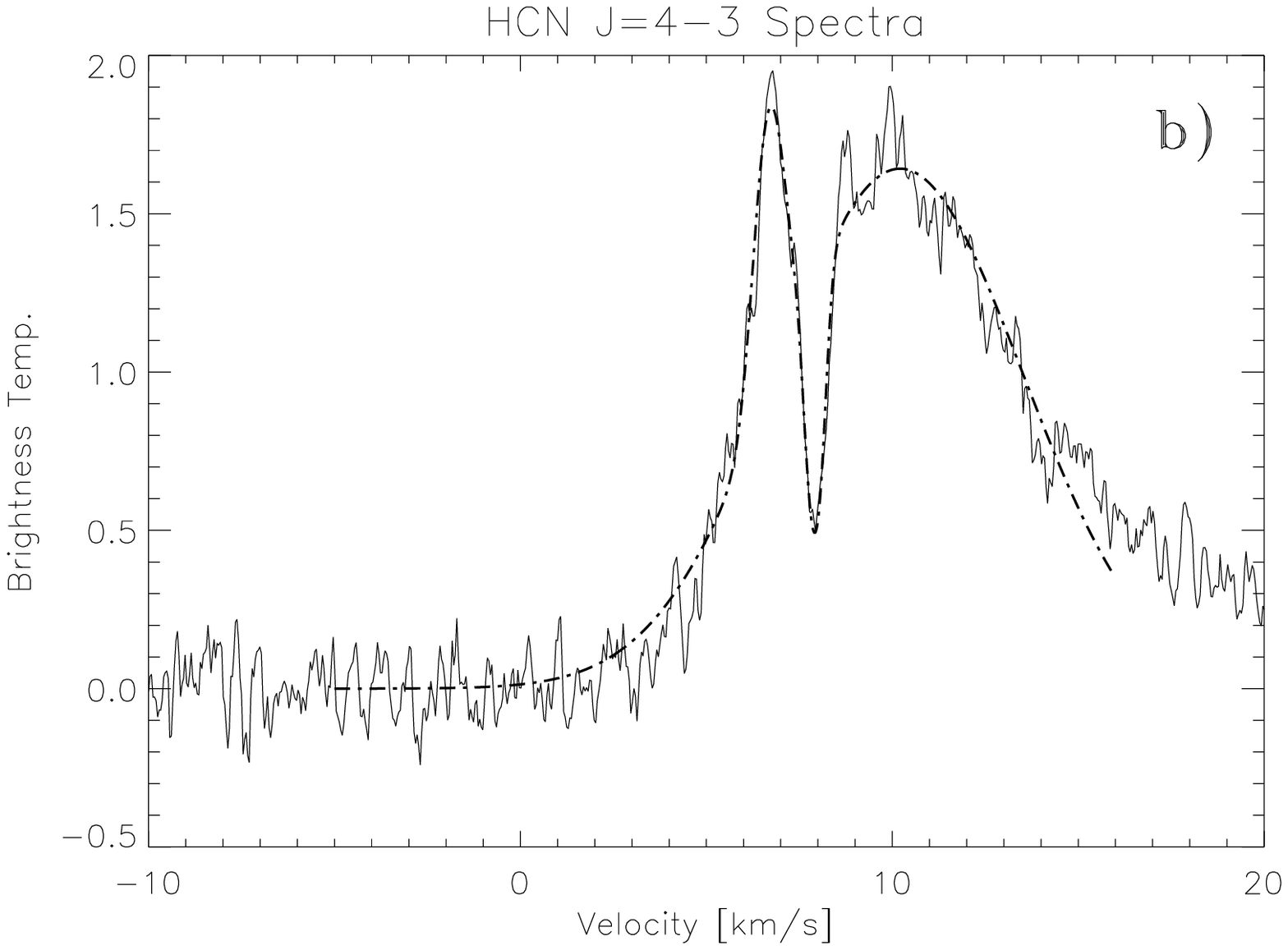}

\epsscale{0.5}
\plotone{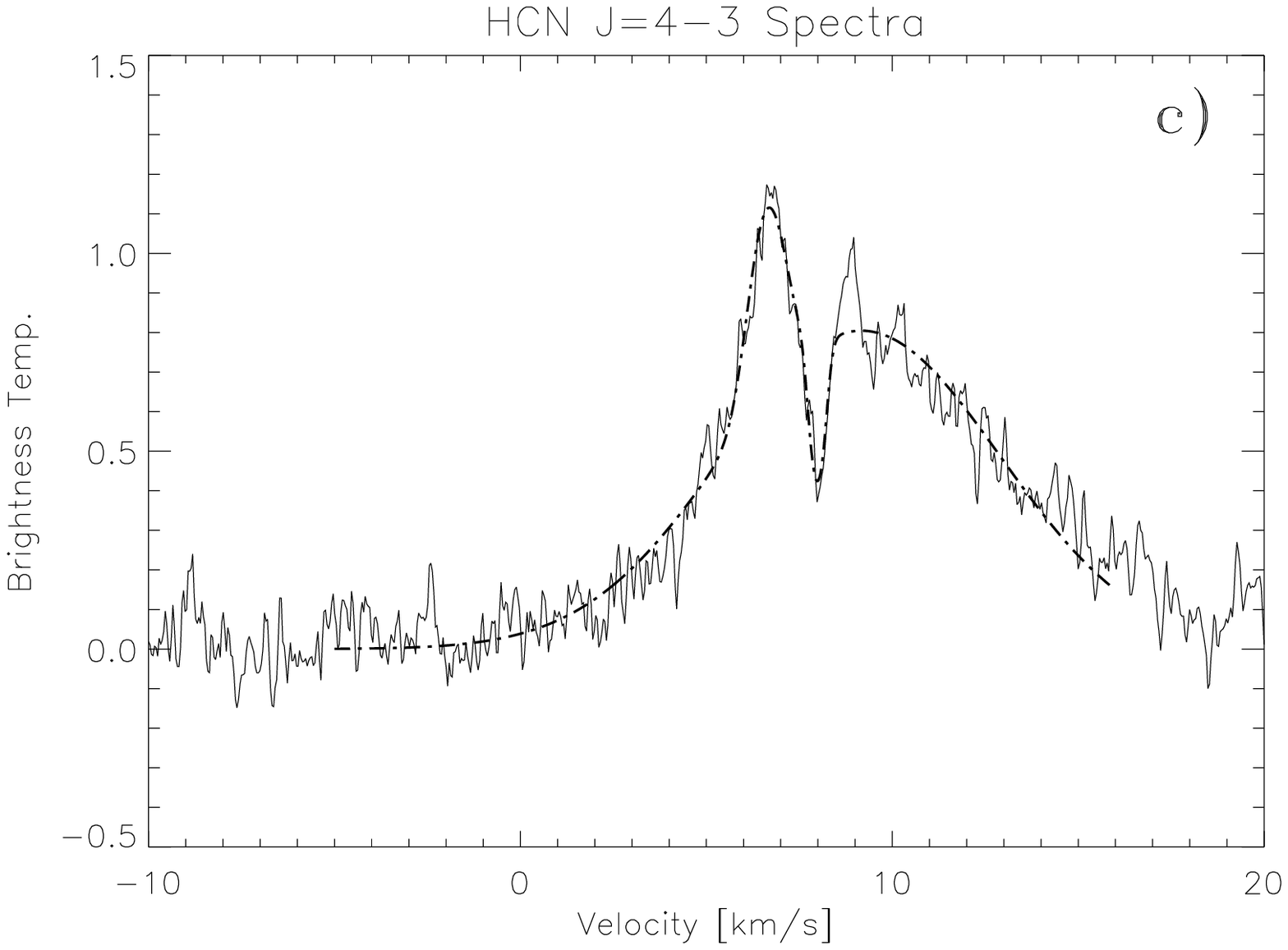}

\caption{$\mathrm{HCN}\: J=4\rightarrow3$ line emission observed at offset
positions ($\alpha_{o}=0$\arcsec, $\delta_{o}=0$\arcsec) (a),
($\alpha_{o}=7$\arcsec, $\delta_{o}=20$\arcsec) (b), and ($\alpha_{o}=15$\arcsec,
$\delta_{o}=30$\arcsec) (c) from the peak continuum flux of 4A.
The dashed curves show the fit of our model, equation (4) in the text.}

\end{figure}

\begin{figure}
\epsscale{0.5}
\plotone{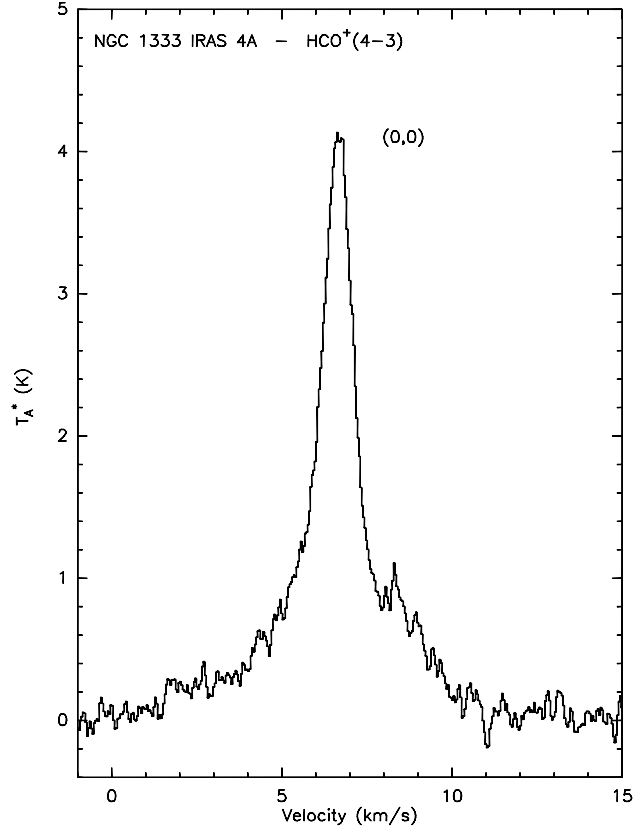}

\caption{$\mathrm{HCO^{+}}\: J=4\rightarrow3$ line emission at an offset position
of ($\alpha_{o}=0$\arcsec, $\delta_{o}=0$\arcsec) from the peak
continuum flux of 4A. These data were used for the evaluation of $\delta v_{\mathrm{los}}$
in our calculation of $B_{\mathrm{pos}}$. We have assumed the ion
will be more closely associated with the magnetic field through the
Lorentz force, and hence the line width observed here should be more
representative of the turbulance/MHD waves that may be distrubing
the embedded field. The line width is calculated to be $\delta v_{\mathrm{los}}\approx1.67\:\mathrm{km/s}$.}

\end{figure}

\clearpage

\begin{deluxetable}{cccccc}

\tabletypesize{\footnotesize}

\tablecaption{Measured Polarization for NGC 1333 IRAS 4\tablenotemark{c}. \label{ta:pol}}

\tablecolumns{6}

\tablewidth{0pt}

\tablehead{

\colhead{$\alpha_{o}$ offset\tablenotemark{b}} & \colhead{$\delta_{o}$ offset\tablenotemark{b}} & \colhead{$p$} & \colhead{$\sigma_{p}$} & \colhead{$\theta_{i}$\tablenotemark{a}} & \colhead{$\sigma_{\theta_{i}}$} \\

\colhead{(arcsec)} & \colhead{(arcsec)} & \colhead{($\%$)} & \colhead{($\%$)} & \colhead{($^{\circ}$)} & \colhead{($^{\circ}$)}

}

\startdata

19.4 & 0.3 & 1.8 & 0.5 & 55.3 & 7.1 \\
13.9 & 21.1 & 4.6 & 1.2 & 61.6 & 8.8 \\
6.1 & 5.0 & 1.5 & 0.5 & 81.9 & 7.7 \\
5.0 & 9.5 & 1.9 & 0.5 & 74.1 & 8.0 \\
1.6 & 10.1 & 1.3 & 0.3 & 73.9 & 6.9 \\ \hline
-8.1 & 6.3 & 1.2 & 0.3 & 43.3 & 15.0 \\
-7.9 & 10.3 & 1.3 & 0.4 & 63.1 & 9.1 \\
-9.8 & 23.3 & 2.3 & 0.8 & 40.5 & 9.8 \\
-10.5 & 27.0 & 2.8 & 0.9 & 39.9 & 5.0 \\
-13.8 & 5.2 & 1.8 & 0.6 & 41.2 & 10.6 \\
-14.0 & 19.3 & 1.9 & 0.3 & 42.0 & 3.5 \\
-13.7 & 24.0 & 1.8 & 0.4 & 52.6 & 8.5 \\
-15.5 & 27.5 & 2.7 & 0.7 & 55.4 & 5.7 \\
-19.7 & 13.6 & 2.5 & 0.7 & 52.4 & 6.0 \\
-18.6 & 19.2 & 1.8 & 0.4 & 55.1 & 7.4 \\
-19.3 & 23.6 & 2.2 & 0.6 & 73.1 & 6.3 \\
-25.0 & 11.4 & 3.4 & 0.9 & 24.2 & 7.0 \\
-25.5 & 15.9 & 3.6 & 1.1 & 30.8 & 8.3 \\
-30.5 & 10.5 & 4.0 & 0.8 & 28.8 & 5.8 \\

\enddata

\tablenotetext{a}{Note $\theta_{i}$ angles describe the orientation of the deduced magnetic field. Angles are measured relative to north and increasing eastward}

\tablenotetext{b}{Offset positions with respect to the peak position of NGC 1333 IRAS 4B}

\tablenotetext{c}{Data given below the solid line describe vectors associated with 4A}

\end{deluxetable}

\clearpage

\begin{deluxetable}{ccccccccccccc}

\tabletypesize{\footnotesize}

\tablecaption{Model Parameters for HCN $J=4\rightarrow3$ Inverse P Cygni Profiles\tablenotemark{b}. \label{ta:spe}}

\rotate

\tablecolumns{13}

\tablewidth{0pt}

\tablehead{

\colhead{$\alpha_{o}$,$ \delta_{o}$\tablenotemark{a}} & \colhead{$J(T_{f})$} & \colhead{$J(T_{r})$} & \colhead{$\tau_{0f}$} & \colhead{$\tau_{0r}$} & \colhead{$V_{f}$} & 
\colhead{$\sigma_{f}$} & \colhead{$J(T_{R})$} & \colhead{$V_{R}$} & \colhead{$\sigma_{R}$} & \colhead{$\tau_{R}$} & \colhead{$V_{r}$} & 
\colhead{$\sigma_{r}$} \\

\colhead{} & \colhead{(K)} & \colhead{(K)} & \colhead{} & \colhead{} & \colhead{(km/s)} & \colhead{(km/s)} & \colhead{(K)} & \colhead{(km/s)} & \colhead{(km/s)} & \colhead{} & \colhead{(km/s)} & \colhead{(km/s)}

}

\startdata

0$\arcsec$, 0$\arcsec$ & 0.30 & 11.00 & 0.49 & 0.15 & 0.95 & 0.36 & 107.14 & -2.86 & 5.44 & 0.01 & 
0.30 & 0.44 \\
7$\arcsec$, 20$\arcsec$ & 0.37 & 4.34 & 0.84 & 0.37 & 0.96 & 0.22 & 128.59 & 10.21 & 3.29 & 0.01 &
0.26 & 0.40 \\
15$\arcsec$, 30$\arcsec$ & 0.75 & 2.50 & 1.15 & 0.42 & 1.03 & 0.18 & 118.57 & 9.17 & 3.72 & 0.01 & 
0.31 & 0.47 \\
 
\enddata

\tablenotetext{a}{Offset positions with respect to the peak position of NGC 1333 IRAS 4A}

\tablenotetext{b}{Variable definitions are given in Section 3.1}

\end{deluxetable}
\end{document}